# Photon trapping in cavity quantum electrodynamics

G. S. Agarwal[1] and Yifu Zhu[2]

[1]*Department of Physics, Oklahoma State University, Stillwater, Oklahoma 74078, USA*
[2]*Department of Physics, Florida International University, Miami, Florida 33199, USA*

We propose and analyze a scheme for photon trapping in an optical resonator coupled with two-level atoms. We show that when the cavity is excited by two identical light fields from two ends of the cavity respectively, the output light from the cavity is suppressed while the intra-cavity light field is near the maximum due to the excitation of the polariton state of the coupled cavity and atom system. We also present methods for the direct probing of the trapped polariton state. The photon trapping is manifested by the destructive interference of the transmitted light and the incident light. Such photon trapping is quite generic and should be observable experimentally in a variety of cavity quantum electrodynamics systems.

Photon confinement and trapping is a current research topic and is important for a variety of fundamental studies and practical applications [1-4]. Considerable research efforts have been spent in exploring new ideas and developing practical techniques to slow down, localize, trap, and store photons in atomic media or photonic structures [1-7]. A well-know example is the Anderson localization in which light can be trapped in a disordered medium through multiple light scatterings [2-3]. Recently, the light trapping and localization have been reported in the nano-plasmonics device and nano-optical structures [6-7]

Here we propose a novel scheme for photon trapping in a cavity quantum electrodynamics (CQED) system [8] and no disordered medium is involved [9]. The CQED system consists of a cavity containing N two-level atoms and being excited by two coherent light fields from two output mirrors [10]. We show that when the two input fields are identical, the photons are coupled into the cavity through the polariton excitation by two input fields, but cannot leak out from the cavity. The output light fields from the cavity are completely suppressed. We derive the photon trapping conditions and present numerical calculations that reveal the detailed characteristics of the photon trapping in the CQED system.

 Fig. 1 shows the schematic diagram for the coupled CQED system that consists of N two-level atoms confined in a single mode cavity and is excited by two input light fields $a_{in}^r$ and $a_{in}^l$ from two ends. The cavity mode couples the atomic transition |2>-|1> (|2> (|1>) is the ground (excited) state of the two-level atoms) with frequency detuning $\Delta_c = \nu_{cav} - \nu_{21}$. The two input fields have the same frequency $\nu_l$ and is coupled into the cavity with the frequency detuning $\delta = \nu_l - \nu_c$. The frequency detuning of the input fields from the atomic transition is $\Delta = \nu_l - \nu_{21}$. We define the collective atomic operators $S^z = \frac{1}{2}\sum_{i=1}^{N}(s_{11}^i - s_{22}^i)$, $S^- = \sum_{i=1}^{N} s_{21}^i$, and $S^+ = \sum_{i=1}^{N} s_{12}^i$ ($s_{11}^i$, $s_{22}^i$, $s_{21}^i$, and $s_{12}^i$ are the atomic operators for the ith atom). The interaction Hamiltonian for the CQED system is

$$H = \delta \cdot a^+ a + \Delta S_z + \{ g a S^+ + i a^+ (\sqrt{2\kappa_r}\, a_{in}^r + \sqrt{2\kappa_l}\, a_{in}^l) + H.C.\}\, . \quad (1)$$

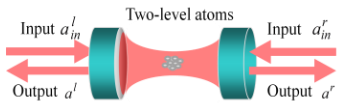

Fig. 1 The CQED system consisting of N two-level atoms confined in the cavity mode and two input light fields.

Here $\hat{a}$ ($\hat{a}^+$) is the annihilation (creation) operator of the cavity photons, $a_{in}^l$ and $a_{in}^r$ are two input fields to the cavity (see Fig. 1), $\kappa_i = \dfrac{T_i}{\tau}$ (i=r or l) is the loss rate of the cavity field on the mirror i ($T_i$ is the mirror transmission and $\tau$ is the photon round trip time inside the cavity), and $g = \mu_{12}\sqrt{\omega_c / 2\hbar \varepsilon_0 V}$ is the cavity-atom coupling coefficients and is assume to be uniform for the N identical atoms inside the cavity. The equations of motion for the CQED system, $\dfrac{d\hat{\rho}}{dt} = -[H, \hat{\rho}] + \hat{L}\hat{\rho}$, are

$$\dot{S}^z = -2\Gamma(S_z + N/2) - i(gaS^+ - ga^+ S^-)\,, \quad (2\text{-}1)$$

$$\dot{S}^+ = -(\Gamma - i\Delta_p)S^+ - 2iga^+ S^z\,, \quad (2\text{-}2)$$

$$\dot{a} = -((\kappa_1 + \kappa_2)/2 + i(\Delta_c - \Delta_p))a - igS^- + \sqrt{\kappa_1/\tau}\, a_{in}^r + \sqrt{\kappa_2/\tau}\, a_{in}^l\,. \quad (2\text{-}3)$$

$2\Gamma$ is the decay rate of the excited state |2>. We consider a symmetric cavity in which $\kappa_1 = \kappa_2 = \kappa$, drop the quantum fluctuation terms, and treat $S^+$, $S^-$, $S^z$, and $a$ as c numbers. In the weak excitation limit ($S^z \approx -N/2$), the steady-state solution of the output light field from the right mirror and the left mirror are

$$a^r = \dfrac{\kappa(a_{in}^r + a_{in}^l)}{\kappa - i\delta + \dfrac{g^2 N}{\Gamma - i\Delta}} - a_{in}^r\,, \quad \text{and} \quad (3\text{-}1)$$

$$a^l = \dfrac{\kappa(a_{in}^r + a_{in}^l)}{\kappa - i\delta + \dfrac{g^2 N}{\Gamma - i\Delta}} - a_{in}^l\,, \quad (3\text{-}2)$$

respectively. If the two input fields are identical, $a_{in}^r = a_{in}^l$, the two output fields are equal, $a^r = a^l$. When $\kappa - i\delta + \frac{g^2 N}{\Gamma - i\Delta} = 2\kappa$, $a^r = a^l = 0$ ( but the intra-cavity light field $a \neq 0$ ), the photons inside the cavity cannot leak out and the photon trapping occurs in the CQED system. The specific trapping conditions are:

$$\frac{\delta}{\Delta} = \frac{\kappa}{\Gamma}, \qquad (4\text{-}1)$$

and

$$\delta\Delta = g^2 N - \kappa\Gamma. \qquad (4\text{-}2)$$

The physics behind the photon trapping is the destructive interference between the transmitted filed and the input light field [10] as depicted in Fig. 2 for the case when the cavity is resonant with the atomic transition.

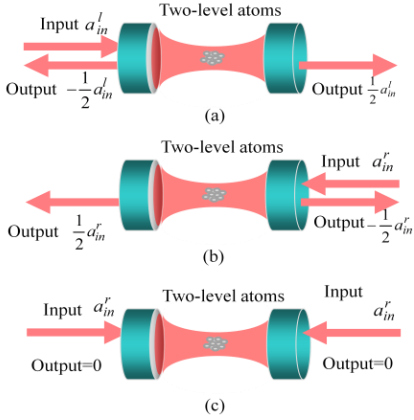

Fig. 2 (a) The CQED system with only the left input field $a_{in}^l$. (b) The CQED system with only the right input field $a_{in}^r$. (c) The CQED system with both input fields $a_{in}^l$ and $a_{in}^r$. With κ=Γ, the coherent addition of (a) and (b) leads to the photon trapping in (c).

In the strong collective coupling limit, $g^2 N > \Gamma\kappa$, the CQED system is resonantly excited when the input field frequency is tuned to the resonant frequency of the polariton states (the normal modes) at $\Delta = \lambda_\pm = \pm g\sqrt{N}$ (the cavity detuning $\Delta_c=0$). If there is only one input field (Fig. 2(a)), the left output field is $a^l = -\frac{1}{2}a_{in}^l$ and out of phase with the input field $a_{in}^l$; the right output field $a^r = \frac{1}{2}a_{in}^l$ and is in phase with the input field $a_{in}^l$. No photon trapping occurs as showed in Fig. 3 for the output light intensities from two ends of the cavity. Fig. 3(a) shows the standard two-peaked spectrum of the CQED system with the peak separation equal to the vacuum Rabi frequency $2g\sqrt{N}$ [11-14]. Fig. 3(b) plots the normalized intra-cavity light intensity $|a|^2 / |a_{in}^l|^2$, (the photonic excitation) and $|b|^2 = |S^+|^2 / N$ (the total atomic excitation). At the peak of the polariton excitation, $|a|^2 + |b|^2 + |a^r|^2 + |a^l|^2 = |a_{in}^l|^2$.

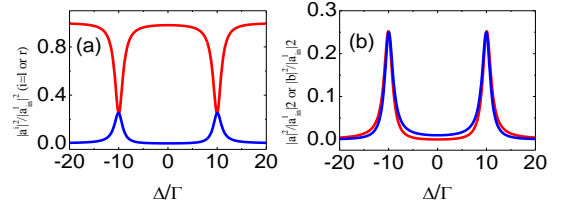

Fig. 3 With only one input field $a_{in}^l \neq 0$ ( $a_{in}^r = 0$ ) and $g\sqrt{N} = 10\Gamma$, (a) the output field intensities $|a^l|^2$ (red line) and $|a^r|^2$ (blue line) normalized to $|a_{in}^l|^2$ versus the input frequency detuning Δ/Γ; (b) the intracavity field intensity $|a|^2$ (red line) and $|b|^2$ (blue line) normalized to $|a_{in}^l|^2$ versus Δ/Γ.

Similarly, if there is only one input field $a^r$ from the right side, the left output field becomes $a^l = \frac{1}{2}a_{in}^r$ and the right output field becomes $a^r = -\frac{1}{2}a_{in}^r$. No photon trapping occurs. However, if both input fields are present and are equal in their phase and amplitude, $a^r = a^l$, the coherent addition of the Fig. 2(a) and Fig. 2(b) leads to the combined output field at the left side $a^l = \frac{1}{2}a_{in}^r - \frac{1}{2}a_{in}^r = 0$ and the right side $a^r = -\frac{1}{2}a_{in}^r + \frac{1}{2}a_{in}^l = 0$. That is, the photons are trapped inside the cavity and cannot leak out from the cavity mirrors. We performed numerical calculation for the photon trapping in the CQED system with system parameters consistent with that reported in earlier CQED experiments [15]. The detailed behavior of the photon trapping is shown in Fig. 4(a) in which the output light intensity $|a^r|^2$ ($=|a^l|^2$) is plotted versus Δ. It shows that at the polariton resonance $\Delta = \pm g\sqrt{N}$, when the photon trapping condition κ=Γ is satisfied (the blue line), the output light is completely suppressed; when the photon trapping condition is not met (κ=0.1Γ, the red line or κ=3Γ, the brown line), the output light intensity is not zero. With $\Delta_c=0$, the spectra are symmetrical and both normal modes exhibit the photon trapping when the input fields are tuned to their resonances at $\Delta = \pm g\sqrt{N}$. The photon trapping also occurs when $\Delta_c \neq 0$ (the cavity is tuned away from the atomic resonance). From the trapping conditions (4-1) and (4-2), one derives $\Delta = \pm\sqrt{\frac{\Gamma g^2 N}{\kappa} - \Gamma^2}$ and $\delta = \pm\sqrt{\frac{\kappa g^2 N}{\Gamma} - \kappa^2}$, which results in $\Delta_c = \pm\{\sqrt{\frac{\Gamma g^2 N}{\kappa} - \Gamma^2} - \sqrt{\frac{\kappa g^2 N}{\Gamma} - \kappa^2}\}$. The resonant frequencies of the polariton states (normal modes) are given by $\lambda_\pm = \frac{\Delta_c}{2} \pm \sqrt{\frac{\Delta_c^2}{4} + g^2 N}$. Then in order for the photon trapping to occur, two conditions must be simultaneously met: first, the input laser detuning must match $\Delta = \lambda_\pm$ such that the polariton state (normal mode) of the CQED system can be excited and the photons are coupled into the cavity mode; second, the same detuning $\Delta$ must also satisfy $\Delta = \pm\sqrt{\frac{\Gamma g^2 N}{\kappa} - \Gamma^2}$. As one example, we consider $g^2 N \gg \Gamma\kappa$, then the photon trapping conditions are $\Delta \approx \pm g\sqrt{N}\sqrt{\frac{\Gamma}{\kappa}}$ and $\delta \approx \pm g\sqrt{N}\sqrt{\frac{\kappa}{\Gamma}}$, which leads to $\Delta_c = \pm g\sqrt{N}(\sqrt{\frac{\Gamma}{\kappa}} - \sqrt{\frac{\kappa}{\Gamma}})$. In an experimental CQED system with a fixed atomic transition frequency, typically, the cavity decay rate κ and the atomic decay rate Γ are fixed, but the cavity frequency detuning $\Delta_c$ can be freely tuned. Then, with $\Delta_c = +g\sqrt{N}(\sqrt{\frac{\Gamma}{\kappa}} - \sqrt{\frac{\kappa}{\Gamma}})$ ($\delta \approx +g\sqrt{N}\sqrt{\frac{\kappa}{\Gamma}}$ and $\Delta \approx +g\sqrt{N}\sqrt{\frac{\Gamma}{\kappa}}$), one derives $\lambda_+ \approx +g\sqrt{N}\sqrt{\frac{\Gamma}{\kappa}}$ and $\lambda_- \approx -g\sqrt{N}\sqrt{\frac{\kappa}{\Gamma}}$. The two trapping conditions for the input light frequency $\Delta$ are simultaneously met only at $\Delta \approx +g\sqrt{N}\sqrt{\frac{\Gamma}{\kappa}} = \lambda_+$, but not $\lambda_-$. Therefore, the photon trapping occurs only at one of the polariton states at $\lambda_+$. Similarly, with $\Delta_c = -g\sqrt{N}(\sqrt{\frac{\Gamma}{\kappa}} - \sqrt{\frac{\kappa}{\Gamma}})$ ($\delta \approx -g\sqrt{N}\sqrt{\frac{\kappa}{\Gamma}}$ and $\Delta \approx -g\sqrt{N}\sqrt{\frac{\Gamma}{\kappa}}$), the resonant frequencies of the polariton states are $\lambda_+ \approx +g\sqrt{N}\sqrt{\frac{\kappa}{\Gamma}}$ and $\lambda_- \approx -g\sqrt{N}\sqrt{\frac{\Gamma}{\kappa}}$. The photon trapping occurs only at the polariton state $\lambda_-$. One example is plotted in Fig. 5 that shows with a detuned cavity ($\Delta_c = 15\Gamma$), the excitation spectrum is asymmetrical, both output light fields $a_{in}^l$ and $a_{in}^r$ are completely suppressed at the left polariton peak $\lambda_- \approx -g\sqrt{N}\sqrt{\frac{\Gamma}{\kappa}} = -5\Gamma$; at the right polariton state $\lambda_+ \approx +g\sqrt{N}\sqrt{\frac{\kappa}{\Gamma}} = 20\Gamma$, the output light field is nonzero.

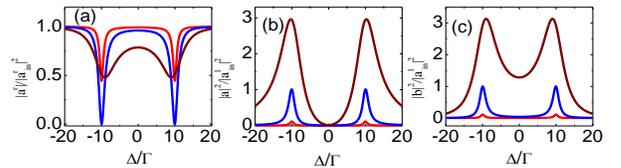

Fig. 4 With two identical input fields $a_{in}^l = a_{in}^r$, $g\sqrt{N} = 10\Gamma$ and $\Delta_c=0$, (a) the output field intensity $|a^l|^2$ and $|a^r|^2$ ($|a^r|^2 = |a^l|^2$) normalized to $|a_{in}^l|^2$ versus the input frequency detuning Δ/Γ; (b) the intracavity field intensity $|a|^2$ and (c) $|b|^2$ versus Δ/Γ. The blue lines correspond to the photon trapping with κ=Γ.

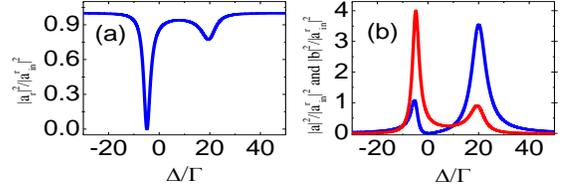

Fig. 5 With two identical input fields $a_{in}^l = a_{in}^r$, $g\sqrt{N} = 10\Gamma$, and κ=4Γ, and the photon trapping condition for the cavity detuning is met at $\Delta_c = g\sqrt{N}(\sqrt{\frac{\Gamma}{\kappa}} - \sqrt{\frac{\kappa}{\Gamma}}) = 15\Gamma$, (a) the output field intensity $|a^l|^2$ and $|a^r|^2$ ($|a^l|^2 = |a^l|^2$) normalized to $|a_{in}^l|^2$ versus Δ/Γ; (b) the intracavity field intensity $|a|^2$ and the atomic excitation $|b|^2$ normalized to $|a_{in}^l|^2$ versus Δ/Γ.

Fig. 5(b) plots the atomic excitation $|b|^2$ (the red line) and the intra-cavity photonic excitation $|a|^2$ versus Δ and shows that the polariton state at $\lambda_- \approx -g\sqrt{N}\sqrt{\frac{\Gamma}{\kappa}} = -5\Gamma$ is largely consisted of the atomic excited state and the polariton state at $\lambda_+ \approx +g\sqrt{N}\sqrt{\frac{\kappa}{\Gamma}} = 20\Gamma$ is largely consisted of the photonic state. The photon trapping occurs to the polariton state dominated by the atomic excitations [16-18].

There is no output light when the photon trapping occurs, but the photon trapping spectral and dynamic properties can be measured by applying a free-space probe laser that couples the polariton state to a 3rd atomic state (see Fig. 6) with the technique demonstrated in ]19]. In Fig. 6(a), the weak probe

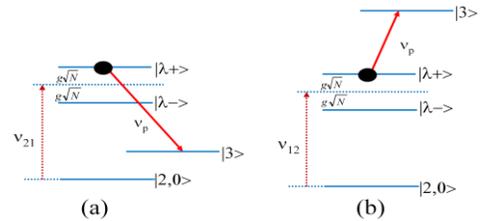

Fig. 6 (a) a Λ-type system (b) a Ladder-type system for probing the trapped |λ+> polariton state.

laser will be amplified when it is tuned to |λ+>-|3> transition but would be unaffected if it is tuned to |λ->-|3> transition. In Fig. 6(b), the fluorescence occurs from the upper state |3> when the probe is tuned to |λ+>-|3> transition but no fluorescence occurs for the probe tuned to |λ->-|3> transition. The probe laser not only can be used to characterize the photon trapping but it can also be used as a control field to explore the photon trapping for possible applications.

The photon trapping in the CQED system can be controlled by the relative phase between the two input fields $a_{in}^l$ and $a_{in}^r$. With the photon trapping conditions (4-1) and (4-2) satisfied, Fig. 7 shows the photon trapping dependence on the phase difference $\Delta\Phi = \varphi_l - \varphi_r$ of the two input fields $a_{in}^l = |a_{in}^l| e^{i\varphi_l}$ and $a_{in}^r = |a_{in}^r| e^{i\varphi_r}$. With $\Delta_c = 0$ (the cavity is resonant with the atomic transition) and $\Delta = g\sqrt{N}$ (the input light is resonant with the polariton state $|\lambda_+\rangle$), Fig. 6(a) and 6(b) plot the output light intensities $|a^l|^2$ and $|a^r|^2$, and $|a|^2$ and $|b|^2$ versus $\Delta\Phi$. It shows that by varying $\Delta\Phi$ from 0 to $\pi$, the output light intensities change from zero to the maximum value 1 while the intra-cavity excitation ($|a|^2$ and $|b|^2$) changes from the maximum to zero. The two output fields are not equal to each other except at $\Delta\Phi = 0$ or $\pi$. Similar behavior can be also observed for the cavity detuned from the atomic resonance as shown in Fig. 6(a') and (b'), in which $\Delta_c = 15\Gamma$ and the input field frequency is tuned to the polariton resonance at $\Delta = -5\Gamma$ (the same parameters as in Fig. 5). The specific technical parameters ($g\sqrt{N} = 10\Gamma$ and $\kappa = \Gamma$ or $\kappa = 4\Gamma$) used for the calculations here can be readily obtained in a reported experimental CQED system with cold Rb atoms [16]. Therefore, the experimental observation of the photon trapping in the CQED system with cold Rb atoms should be feasible.

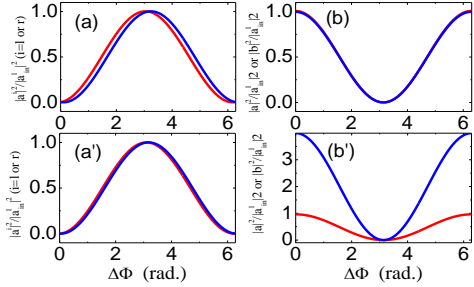

Fig. 7 With two input fields $a_{in}^l = |a_{in}^l| e^{i\varphi_l}$ and $a_{in}^r = |a_{in}^r| e^{i\varphi_r}$, (a) the normalized output field intensity $|a^l|^2$ (the red line) and $|a^r|^2$ (the blue line) and (b) the normalized intracavity field intensity $|a|^2$ (the red line) and the atomic excitation $|b|^2$ (the blue line, nearly overlapped with the red line) versus $\Delta\Phi = \varphi_l - \varphi_r$. $\Delta = g\sqrt{N} = 10\Gamma$, and other parameters are the same as that in Fig. 4. With the cavity detuning $\Delta_c = 15\Gamma$ and $\Delta = -5\Gamma$ (other parameters are the same as that in Fig. 5), (a') plots the normalized $|a^l|^2$ (the red line) and $|a^r|^2$ (the blue line), and (b') plots the normalized $|a|^2$ (the red line) and $|b|^2$ (the blue line) versus $\Delta\Phi$.

In conclusion, we have shown that in a multiatom CQED system coupled by two identical laser fields from the two output ends of the cavity, photons can be trapped inside the cavity when the polariton state is excited. Such photon trapping occurs for the two symmetrically located polariton states when the cavity is resonant with the atomic transition. If the cavity is detuned from the atomic transition, the photon trapping can be observed in one of the two polariton states. The photon trapping proposed here is a general physical phenomenon induced by the destructive interference between the transmitted light field and the input light field, and should be observable in a variety of experimental CQED systems [20-24]. Particularly, it will be interesting to explore the photon trapping in a hybrid ferromagnetic magnons and microwave cavity system [24] in which the Kittel mode frequency (corresponds to the atomic transition frequency $\nu_{12}$ here) can be continuously tuned, thus it provides an alternative way to match the photon trapping conditions.

### Acknowledgement

Y. Zhu acknowledges support from the National Science Foundation under Grant No. 1205565.